 \documentclass[preprint]{aastex}

\newcommand{\etal}{{\em et al.\,}}
\newcommand{\micr}{$\,\mu$m}
\newcommand{\brg}{Br$\gamma$}
\newcommand{\water}{$\rm H_{2}O$}
\newcommand{\ewcoa}{$W_{2.29}$}
\newcommand{\ewcob}{$W_{2.32}$}
\newcommand{\ewcoc}{$W_{2.34}$}
\newcommand{\ewcoh}{$W_{1.62}$}
\newcommand{\ewsi}{$W_{1.59}$}
\newcommand{\ewna}{$W_{\rm Na}$}
\newcommand{\ewca}{$W_{\rm Ca}$}
\newcommand{\ewfea}{$W_{\rm Fe1}$}
\newcommand{\ewfeb}{$W_{\rm Fe2}$}
\newcommand{\ewmg}{$W_{\rm Mg}$}
\newcommand{\cohsi}{${\rm log}\,(W_{1.62}/W_{1.59})$}
\newcommand{\cohcok}{${\rm log}\,(W_{1.62}/W_{2.29})$}
\newcommand{\teff}{$T_{\rm eff}$}

\slugcomment{Accepted for publication in the {\em Astronomical Journal}}
\shorttitle{3D $K$-band Stellar Library}
\shortauthors{F\"orster Schreiber}

\begin{document}

\title{Moderate resolution near-infrared spectroscopy of cool stars: \\
       a new $K$-band library}

\author{N. M. F\"orster Schreiber} 
\affil{Max-Planck-Institut f\"ur extraterrestrische Physik, 
       Postfach 1312, D-85741 Garching, Germany \\
       and \\
       CEA/DSM/DAPNIA/Service d'Astrophysique,
       C.E. Saclay, F-91191 Gif sur Yvette CEDEX, France
       \email{forster@discovery.saclay.cea.fr}}

\begin{abstract}
I present an atlas of near-infrared $K$-band spectra of 31 late-type giants
and supergiants, and two carbon stars.  The spectra were obtained at resolving
powers of 830 and 2000, and have a signal-to-noise ratio $\gtrsim 100$.
These data are complemented with results from similar existing libraries 
in both $K$- and $H$-band, and are used to identify various tools useful
for stellar population studies at moderate resolution.  
I focus on several of the most prominent absorption features and
(1) investigate the effects of spectral resolution on measurements of
their equivalent width (EW),
(2) examine the variations with stellar parameters of the EWs, and
(3) construct composite indices as indicators of stellar parameters 
and of the contribution from excess continuum sources commonly found
in star-forming and AGN galaxies.  
Among the features considered, the $^{12}$CO\,(2,0) and $^{12}$CO\,(6,3)
bandheads together with the \ion{Si}{1} 1.59\micr\ feature, first proposed
by Oliva, Origlia, and coworkers, constitute the best diagnostic set for
stellar spectral classification and for constraining the excess continuum
emission.  The \ion{Ca}{1} 2.26\micr\ and \ion{Mg}{1} 2.28\micr\ features
offer alternatives in the $K$-band to the $^{12}$CO\,(6,3) bandhead and
\ion{Si}{1} feature.
\end{abstract}


\section{Introduction}  \label{Sect-Intro}

Among their many applications, libraries of stellar spectra are particularly 
useful in population and evolutionary synthesis of stellar clusters and
galaxies.  Extensive stellar databases now exist in the ultraviolet and 
optical regimes, and have been applied to numerous studies of Galactic and
extragalactic sources \citep[{\em e.g. \rm}][]{Wor94, Lei96}.
These are, however, of limited usefulness for obscured systems.
The near-infrared regime ($\rm \lambda = 1 - 2.5~\mu m$) offers an
alternative opportunity to probe obscured sources since extinction 
effects are much less important at these wavelengths ({\em e.g.} 
$A_{\rm 2.2\,\mu m} \approx 0.1\,A_{V}$).  In addition, cool
($\sim 3000 - 6000~{\rm K}$) stellar populations are best studied 
in the near-infrared because their spectral energy distributions
peak near 1\micr.  

Since the pioneering work of \citet{Joh70}, and especially
in the past decade due to improvements in instrumentation and detector 
sensitivity, many investigators have conducted near-infrared spectroscopic
studies of stars, covering overall wide ranges in stellar types and spectral
resolution (see {\em e.g.} \citealt*{OMO93}, \citealt{WH97}, and
\citealt{Mey98} for reviews).
A number of atlases at moderate resolving powers in the $H$- and $K$-band
($\rm \lambda = 1.45-1.85~\mu m$ and $\rm \lambda = 1.9-2.5~\mu m$
respectively) have been assembled, sampling various types of cool stars
more systematically, and intended for spectral classification and 
population synthesis purposes.  The first such library was provided by 
\citet[hereafter KH86]{KH86}, and was followed notably by the work of
\citet{LRV92}, \citet[OMO93]{OMO93}, \citet{Ali95}, \citet*[DBJ96]{DBJ96},
\citet{Ram97}, \citet{WH97}, \citet{Lan98}, and \citet{Mey98}.
Some of these authors interpreted their data together with
theoretical stellar atmosphere models as well as with observations of
template composite populations to better understand the dependence
of key spectral features on stellar parameters ({\em e.g.} OMO93;
\citealt{Ali95, OFFO97, OO98}).  Lan\c{c}on and her co-workers
implemented their empirical library in evolutionary synthesis
codes to model stellar clusters and galaxies \citep{LRV96, Lan99}.

Despite these contributions, additional efforts would be beneficial to
improve the quality and consistency of quantitative diagnostic tools 
suitable for stellar population studies.  Specific problems include
the disparity in spectral resolution of the various atlases, and 
the different definitions of spectroscopic indices used to quantify
the strength of the stellar absorption features.  Some studies targeted
specific regions of the Hertzsprung-Russell diagram while others aimed
at a larger coverage but with coarser sampling.  Altogether, the existing
libraries relevant for cool stars constitute a rather heterogeneous data set.

In this context, I have assembled a new near-infrared stellar library
obtained with the Max-Planck-Institut f\"ur extraterrestrische Physik
(MPE) 3D instrument \citep{Wei96}.  It is intended at widening
existing libraries primarily for applications to extragalactic studies.
The 3D atlas includes normal late-type giants and supergiants as well as
asymptotic giant branch --- or AGB --- stars, which usually dominate the
near-infrared continuum emission of galaxies \citep[{\em e.g. \rm}][]{BC93}.
In this paper, I present $K$-band spectra at 
$R \equiv \lambda/\Delta\lambda \approx 830$ and 2000 with
signal-to-noise ratio $\rm S/N \gtrsim 100$ of 31 giants and supergiants
mostly with near-solar abundances, and of two representative carbon stars.
In order to obtain a larger and, ultimately, an homogeneous data set,
the 3D sample is augmented with data from similar libraries in the $H$-
and $K$-band selected from the literature.  I focus on several of the most
prominent absorption features, investigate the effects of spectral resolution
on measurements of their equivalent widths (EWs), and compile consistently the 
results from the different data sets.  I use the data to identify 
diagnostics of the stellar population composition and of excess 
continuum emission (such as due to hot dust heated by OB stars or
by an AGN).  Additional spectra in the $H$-band and of AGB stars
will be presented and analyzed in future contributions.

The paper is organized as follows.
I describe the selection of the program stars, the observations, and the
data reduction in section~\ref{Sect-Obs}.
In section~\ref{Sect-Res}, I present the results, investigate the effects of
spectral resolution on the EWs, and augment the 3D data set with EWs obtained
from selected existing atlases.
In section~\ref{Sect-Indic}, I discuss the variations of the EWs with stellar
parameters and explore various composite spectroscopic indices.
The paper is summarized in section~\ref{Sect-conclu}.

\section{Selection, observations and data reduction of the 3D sample} 
         \label{Sect-Obs}

\subsection{Selection of program stars}  \label{Sub-cat}

Table~\ref{tab-cat} lists the program stars, their spectral types, and their 
metallicities from three literature sources \citep{McW90, Tay91, Cay92}.
The sample includes 25 giants (luminosity class III) and six supergiants
(luminosity classes I and II) of late-G, K, and M types.  They were
selected from the list of optical spectral standards and well-classified
stars of \citet{KM89}, and two additional stars were taken from \citet{Cay92}.
The spectral types are given in the revised MK system \citep{Kee87} and 
were cross-checked with \citet{Kee88}, \citet{Gar94}, \citet{Kee80}, and 
\citet{MK73}.  Stars belonging to double or multiple systems were excluded 
in the selection.  Except for a few giants with $\rm [Fe/H] < -0.3$, none 
of the stars are known to have strong abundance anomalies.
In addition, one N-type and one early-R type carbon stars drawn from
the sample of \citet{Yam72} were observed.  The spectral types assigned
by \citeauthor{Yam72} in the C-classification scheme of \citet{Kee41}
are given, as well as those he quotes in the R- and N-system used in 
the Henry Draper Catalogue \citep[{\em e.g. \rm}][]{Sha28}.

\subsection{Observations}  \label{Sub-Obs}

The stars were observed using the MPE near-infrared integral field spectrometer
3D \citep{Wei96}.  A first set of spectra was obtained at the 3.5~m
telescope in Calar Alto, Spain, on 1995 January 12, 18, and 20.  A second 
set was collected at the 4.2~m William-Herschel-Telescope in La Palma,
Canary Islands, on 1996 January 2 and 9.  The observations were completed
at the 2.2~m ESO telescope in La Silla, Chile, between 1996 April 10 and 20.
3D images a square field of view at
$\rm 0.3^{\prime\prime} - 0.5^{\prime\prime}$ per pixel and provides
simultaneously the $H$- or $K$-band spectrum of each spatial pixel.
The focal plane is sliced in 16 parallel
``slits'' whose light is dispersed in wavelength by a grism onto
a 256$\times$256~HgCdTe~NICMOS~3 array.  The data are rearranged in a
three-dimensional cube during data processing.  The intrinsic spectral 
sampling of 3D is with pixels of size $\rm \lambda/R$; Nyquist-sampled 
spectra are achieved by dithering the sampling by half a pixel on alternate
data sets which are interleaved in wavelength (``merged'') during data 
reduction.  The Calar Alto and La Palma data were obtained at $R \approx 830$ 
in the range $\rm \lambda = 1.9 - 2.4~\mu m$ and the La Silla data, at 
$R \approx 2000$ between 2.18\micr\ and 2.45\micr.

Each star was observed in two sequences of several exposures,
each sequence taken with spectral sampling shifted by half a pixel
with respect to the other.  Two similar sequences were obtained at
off-source positions at most 1\arcmin\ away from the stars.  To
further minimize the effects of variable background emission, the on- 
and off-source sequences were alternated on timescales of less than
10 minutes.  The integration time and number of individual exposures
were dictated by the brightness of each star and by the requirement
of achieving a $\rm S/N \gtrsim 100$ per wavelength channel.
For atmospheric calibration, B, A, F, and early-G type stars
were observed each night, before and after the program stars.
Reference stars close to the library stars were chosen to optimize
the correction for differential airmass and transmission spectrum.
Each night, images were taken of the observatory's dome and of a 
glowing Nernst rod for spatial and spectral flat-fielding, as well as
of a neon discharge lamp for wavelength calibration.
Table~\ref{tab-Obs} gives the log of the observations.

\subsection{Data reduction}  \label{Sub-datared}

The data reduction was carried out using the 3D data analysis package, 
within the GIPSY environment \citep{vdH92}.  The basic steps were performed
following standard procedures as described in \citet{Wei96}.
These include correction for the non-linear signal response of the detector,
averaging of single-frame exposures, dark current and background subtraction,
flat-fielding, wavelength calibration, spectral redistribution onto a regular
grid and merging, rearrangement of the data in a three-dimensional cube,
and correction for bad and hot pixels.  Residuals due to spatial and
temporal variations in the background emission level are $\lesssim 1\%$.

The atmospheric transmission was corrected for with the help of
the reference stars data reduced in the manner described above.  
The calibration spectra extracted from the data cubes were divided
by a black-body curve of appropriate temperature given the spectral
type of each reference star.  The intrinsic absorption features of the
late-F and early-G dwarf calibrators were corrected for by division with
the normalized spectrum of the \ion{G3}{5} star from the KH86 atlas,
convolved with a gaussian profile to the resolution of the 3D data.
This template star has similar line strengths as the calibrators except
for the \brg\ absorption at 2.166\micr, which was removed by linear
interpolation.  The B- and A-type calibrators are expected to be 
featureless at $R \sim 830 - 2000$ apart from \brg\ absorption, 
again removed by linear interpolation.
The exception is the \ion{A8}{1}b calibrator used for
HD\,78647, which likely has absorption from the Pfund series 
of hydrogen longwards of 2.32\micr\ (see \citealt{WH97} who detected
these lines in A through early-F supergiants at $R \sim 3000$).
In the final spectrum of HD\,78647, corresponding residuals are 
however not obvious.

The final spectra of the program stars were extracted from
the data cubes after division by the calibration spectra.
The residuals from intrinsic features of the reference stars are $\leq 1\%$.
Those due to spatial and temporal variations of atmospheric transmission
could be further reduced using synthetic spectra of the differential 
transmission generated with the program ATRAN \citep{Lor92} and by 
adjusting the zenithal distances and water vapour column densities
assigned to the library and reference stars.
Telluric features do not persist to more than 1\% in most of
the band, and 5\% around 2.0\micr\ (10\% for HD\,25025).

At the higher spectral resolution for the La Silla observations, 
and more importantly for the 
$\rm 0.3^{\prime\prime}\,pixel^{-1}$ scale used for several stars,
a complex high-frequency interference pattern appeared in the 
data, due to multiple reflections at the surface of the detector array.
The frequency and position angle of the fringes varied with position 
on the array and with wavelength sampling, and differed significantly
for defocussed observations ({\em e.g.} in the dome images).  This pattern
multiplicatively modulated the observed counts with a typical peak-to-peak
amplitude of $5\% - 10\%$.  The similar nature of the program and reference
sources as well as of the technical conditions under which they were observed
(telescope pointing close to the zenith, same focus, source positioned
at the center of the array) resulted in nearly identical interference
patterns in the respective data sets.  The division of the library stars
data by the calibration spectra cancelled the fringes to better than 1\% 
except for HD\,82668, where they remained at the 3\% level at short 
wavelengths.

The final uncertainties in spectral flux distribution of the reduced data
are estimated to be less than $3\% - 4\%$ at $\rm \lambda > 2~\mu m$, 
and $\leq 10\%$ near and shortwards of 2\micr.  They include possible
systematic errors which may occur in the background subtraction, in the 
merging of the wavelength-shifted data sets, in the correction for telluric
transmission and intrinsic features of the reference stars, and residuals
from the interference pattern in the higher resolution data.
As an indication of the quality of the final spectra, an
effective S/N ratio was evaluated on the reduced data between
2.242\micr\ and 2.258\micr.  It was taken as the inverse of the 
standard deviation about the mean flux after division by a straight line 
fitted over this range; the values are reported in table~\ref{tab-Obs}.
This estimated S/N ratio is intended at providing a measure of the effective,
resultant noise over a relatively line-free portion of the fully reduced
spectra, near the various absorption features discussed in this paper.
For the N-type carbon star HD\,92055, it is however affected by the
presence of significant intrinsic absorption features in the range 
considered (see section~\ref{Sub-spec} below).

\section{Results}  \label{Sect-Res}

\subsection{3D spectra}  \label{Sub-spec}

The spectra are plotted in figures~\ref{fig-SpecLR} and \ref{fig-SpecHR}, 
sorted according to luminosity class and advancing spectral type.  They are
normalized to unity in the interval $\rm \lambda = 2.2875 - 2.2910~\mu m$
and presented on a linear flux scale, vertically shifted by equal intervals
of 0.5 starting at the bottom.  The considerable continuum structure is
most obvious in the higher resolution spectra of figure~\ref{fig-SpecHR}.
The variation in strength of most features with spectral type and luminosity
class is also apparent.  The strongest features longwards of 2.15\micr\
for the normal giants and supergiants were identified by comparing their
locations with those given in KH86.  As shown by \citet{WH96} based on
$R \geq 45000$ spectra, the atomic features result from the blending of
lines from multiple species (see section~\ref{Sub-TeffLC}).  
The nomenclature of KH86 is adopted here to keep with the convention
established so far.  \citet{WH96} also identified numerous lines of the
CN red system shortwards of 2.29\micr\ and, in the latest M-types, of \water\
as well.  At moderate resolution, the blending of these lines produce the
low-level ``grass-like'' continuum structure \citep[see also][]{WH97}.

For the two carbon stars, the CO first overtone bandheads are 
easily recognized as the most prominent features in the 3D spectra.
Figure~\ref{fig-C-III} compares the spectra of the carbon stars 
with those of normal giants of similar effective temperature (see 
section~\ref{Sub-comp} below). 
HD\,92055 (C6,3 -- N2) exhibits similar CO bandhead strengths as 
HD\,80431 (\ion{M4}{3}) but substantially more continuum structure at 
shorter wavelengths.  In the $K$-band spectrum of N-type carbon stars,
believed to lie on the thermally-pulsing AGB 
\citep[see][and references therein]{Wal98},
lines from the CN red system and perhaps some lines of the $\rm C_{2}$
Phillips system become conspicuous, hiding less prominent
atomic features \citep[{\em e.g. \rm}][]{WH96, Lan98}.
On the other hand, early-R stars are presumably in earlier evolutionary
stages and their atmospheres seem to behave as those of normal K giants,
but their nature remains unknown \citep[{\em e.g. \rm}][]{Wal98}.
The spectrum of HD\,113801 (C2,1 -- R0) is 
almost identical to that of its temperature counterpart HD\,107328
(\ion{K0.5}{3}b), suggesting common dominant sources of opacity for
the strongest features, with carbon star features too weak to be
detected at moderate resolution.

The quality of the 3D data is illustrated in figure~\ref{fig-3D-KH}, which
compares several of the higher-resolution spectra with those for similar
spectral types from KH86, convolved to $R \sim 2000$ with a gaussian profile.
The original KH86 spectra have $R \sim 3000$ and $\rm S/N \sim 100 - 1000$,
and are ratioed with \ion{A0}{5} stars.  In order to recover the original
energy distributions, they were multiplied by a power-law continuum 
$\propto \lambda^{-3.94}$ which is the closest representation of an
A0 dwarf (\citealt{Kur79}, as quoted by \citealt{McG87}).
In all cases, the 3D and KH86 data compare remarkably well.

\subsection{Definition of equivalent widths}  \label{Sub-EWs}

The strength of absorption features was quantified by the
equivalent width (EW), defined as:
\begin{equation}
W_{\lambda} = \int^{\lambda_{2}}_{\lambda_{1}}\,
(1 - f_{\lambda})\,{\rm d}\lambda ,
\label{eq-EW}
\end{equation}
where $f_{\lambda}$ is the spectrum normalized to a flat continuum
slope while $\lambda_{1}$ and $\lambda_{2}$ are the integration limits,
and $\rm d\lambda$ is expressed in \AA.
EWs are independent of extinction since the effects of continuous opacity
of interstellar dust grains are cancelled by the continuum normalization.
The EWs were measured for the features involving the \ion{Na}{1} doublet
centered at 2.2076\micr\ and the \ion{Ca}{1} multiplet centered at 2.2636\micr,
and for the $^{12}$CO\,(2,0), $^{12}$CO\,(3,1), and $^{13}$CO\,(2,0) first 
overtone bandheads at 2.2935\micr, 2.3227\micr, and 2.3448\micr\ respectively
(hereafter \ewna, \ewca, \ewcoa, \ewcob, and \ewcoc).  In addition,
the EWs of the features involving the \ion{Fe}{1} lines at 2.2263\micr\ and
2.2387\micr, and the \ion{Mg}{1} line at 2.2814\micr\ (\ewfea, \ewfeb, \ewmg)
were measured in the higher resolution spectra.

To avoid complications if the continuum shape is distorted by broad
absorption features, such as those due to \water\ around 1.9\micr\ and
2.7\micr\ in the coolest stars, or affected by extinction and contributions
from emission sources such as hot dust, the local continuum
was preferred for most features.  
The normalizing continuum for each atomic feature was 
obtained by linear interpolation between adjacent regions 
$\rm 0.002 - 0.003~\mu m$ wide on each side of the absorption.  Since 
there is no line-free continuum redwards of 2.3\micr\ in the 3D spectra, the
normalizing continuum for the CO bandheads was defined at short wavelengths
only.  For $^{12}$CO\,(2,0), it was taken as the average in the clear band
around 2.29\micr.  Although the continuum slope is thereby not constrained,
this does not introduce large uncertainties in \ewcoa\ because the feature
bandpass is narrow and close enough to the continuum interval.  In the case of
$^{12}$CO\,(3,1) and $^{13}$CO\,(2,0), a power-law was fitted to featureless
regions between 2.21\micr\ and 2.29\micr\ to better extrapolate over the 
bandheads.

The narrow bands used in the computation of the EWs are reported in 
table~\ref{tab-Def}.  The integration limits for the \ion{Na}{1} and
\ion{Ca}{1} features and the CO bandheads, as well as the continuum definition
for $^{12}$CO\,(2,0) are as in KH86.  The definition of \ewcoa\ is also
the same as used by OMO93.  In view of the compilation of selected data 
sets from the literature below, the definition of the local continuum 
and bandpass edges adopted by OMO93 to measure the EW of the 
$^{12}$CO\,(6,3) second overtone bandhead at 1.6187\micr\ and the 
\ion{Si}{1} feature at 1.5892\micr\ (\ewcoh\ and \ewsi) are also listed.

\subsection{Effects of spectral resolution}   \label{Sub-resolution}

In order to provide a means for comparing results from data obtained at
different spectral resolution or affected by velocity broadening, the 
variations of the EWs defined above with instrumental resolution were
investigated.  The \ion{Mg}{1} and \ion{Fe}{1} features were excluded 
from this analysis because their weakness hampers reliable measurements
at $R \lesssim 1000$.  The instrumental profiles were represented by 
gaussian functions.

Grids of spectra convolved with gaussians corresponding
to velocity dispersions $\sigma$ ranging from $\rm 0~km\,s^{-1}$ to
$\rm 400~km\,s^{-1}$, in steps of $\rm 10~km\,s^{-1}$, were generated
using the KH86 spectra ($R \sim 3000$) for types later than K0, and 
the 3D spectra at $R \approx 2000$.
The variations of the EWs with $\sigma$ did not depend
on the chosen templates and were best fitted by second order polynomials.
Velocity dispersions $\rm {\sigma} < 60~km\,s^{-1}$ for the CO bandheads,
and $\rm {\sigma} < 20~km\,s^{-1}$ for the \ion{Na}{1} and \ion{Ca}{1}
features produced no measurable effect.  The scatter around the fits
was less than $\pm 5\%$ for $\rm {\sigma} \leq 200~km\,s^{-1}$ but increased
for larger $\sigma$, corresponding to $R \lesssim 600$.  Since at such 
resolving powers, the resolution element $\rm \lambda/R$ becomes comparable
to the integration intervals ${\lambda}_{2} - {\lambda}_{1}$,
$\rm \sigma \leq 200~km\,s^{-1}$ is considered as the range of validity
for the empirical corrections.

\citet{OOKM95} used this approach to derive corrections for \ewsi, \ewcoh,
and \ewcoa\ measured in velocity broadened spectra.  Their correction for
\ewcoa\ obtained using templates at $R \sim 2500$ is given by a linear
relationship for $\rm {\sigma} \geq 60~km\,s^{-1}$.  Within the dispersion
of the results, this line approximates well the fit obtained here in the
range $\rm {\sigma} = 60 - 200~km\,s^{-1}$ and is adopted for consistency with
previous work.  The empirical relationships including those for the $H$-band 
features from \citeauthor{OOKM95} are the following, where 
$W_{\lambda}^{\rm obs}$ denotes the observed EW. 
For \ewsi, \ewcoh, and \ewcoa,
\begin{eqnarray}
W_{\lambda} = W_{\lambda}^{\rm obs}\,\left[1 + 
                                     a_{\lambda}\,(\sigma - 60)\right] \\
{\rm for}~60~{\rm km\,s^{-1}} \leq \sigma \leq 200~{\rm km\,s^{-1}} \nonumber
\label{eq-res1}
\end{eqnarray}
where 
$a_{1.59} = 8.70 \times 10^{-4}$, 
$a_{1.62} = 7.10 \times 10^{-4}$, and
$a_{2.29} = 8.75 \times 10^{-4}$; 
for \ewcob\ and \ewcoc,
\begin{eqnarray}
W_{\lambda} = W_{\lambda}^{\rm obs}\,\left[1 + 
        a_{\lambda}\,(\sigma - 60) + b_{\lambda}\,{(\sigma - 60)}^2\right] \\
{\rm for}~60~{\rm km\,s^{-1}} \leq \sigma \leq 200~{\rm km\,s^{-1}} \nonumber
\label{eq-res2}
\end{eqnarray}
where 
$a_{2.32} = 4.38 \times 10^{-4}$,
$b_{2.32} = 3.52 \times 10^{-6}$,
$a_{2.34} = 1.32 \times 10^{-4}$, and
$b_{2.34} = 2.61 \times 10^{-6}$;
and for \ewna\ and \ewca,
\begin{eqnarray}
W_{\lambda} = W_{\lambda}^{\rm obs}\,\left[1 + 
       a_{\lambda}\,(\sigma - 20) + b_{\lambda}\,{(\sigma - 20)}^2\right]   \\
{\rm for}~20~{\rm km\,s^{-1}} \leq \sigma \leq 200~{\rm km\,s^{-1}} \nonumber
\label{eq-res3}
\end{eqnarray}
where 
$a_{\rm Na} = 1.67 \times 10^{-3}$,
$b_{\rm Na} = 4.43 \times 10^{-6}$,
$a_{\rm Ca} = 2.33 \times 10^{-4}$, and
$b_{\rm Ca} = 6.24 \times 10^{-6}$.
The corrections for $R \sim 1000$ amount to $\approx 5\%$ or less except
for \ewca\ (10\%) and \ewna\ (22\%).  At $R \sim 600$, they are
$\approx 13\%$ or less, 27\%, and 48\% respectively.

\subsection{Compilation of data from 3D and additional selected atlases}  
            \label{Sub-comp}  

The EWs measured on the 3D spectra, corrected for resolution where 
appropriate, are listed in table~\ref{tab-EWs}.  
The atomic features were not measured for the carbon stars.
The atlases of KH86, OMO93, and DBJ96 were chosen to complement the
3D library, providing a larger database.  These data sets
include $H$- and $K$-band spectra at $R \sim 1500 - 3000$.
The EWs were measured on the digitally available spectra of
DBJ96 and KH86; the EWs given in OMO93 were taken directly because
measured according to the definitions adopted here.  Corrections for
instrumental resolution were applied where appropriate.  
Formal uncertainties were derived from the rms noise in the spectra
and propagating this uncertainty through Eq.\,\ref{eq-EW};
average values ranged between 0.1~\AA\ and 0.2~\AA.  The continuum
placement introduces larger uncertainties which were estimated
by varying slightly the continuum-defining intervals, resulting
in about $1\% - 2\%$ error in the continuum level.
The final measurement uncertainties adopted are $\rm \pm 0.5$~\AA\
for the atomic features and \ewcoh, $\rm \pm 0.8$~\AA\ for \ewcoa,
and $\rm \pm 1.0$~\AA\ for \ewcob\ and \ewcoc.  

The variations of the EWs with stellar effective temperature (\teff) are shown 
in figure~\ref{fig-EWs}.  The spectral types were converted to temperatures
using the calibrations for late-type stars of \citet{SK82}.
For the carbon stars, individual determinations were adopted:
$T_{\rm eff} = 4700~{\rm K}$ for HD\,113801 \citep{Dom84} and 
3080~K for HD\,92055 \citep{Ohn96}. 
The EWs for main-sequence stars from KH86 and DBJ96 are included for
comparison purposes.  The data obtained from the various references are 
in excellent agreement.  No systematic deviation of a particular data set
is seen and for the few stars common to two or more samples, the EWs agree
within the measurement uncertainties.

\section{Discussion}  \label{Sect-Indic}

\subsection{Variations of selected features with stellar parameters}
            \label{Sub-TeffLC}

As can be seen from figure~\ref{fig-EWs}, the absorption features of interest
in cool, evolved stars vary strongly with \teff\ over the range of spectral
types considered, except for \ion{Si}{1} and \ion{Mg}{1}.  
The first overtone CO bandheads show the long recognized trend with
luminosity class for $T_{\rm eff} \leq 4500~{\rm K}$, supergiants
exhibiting deeper bandheads than giants of the same temperature.
A similar trend is seen below 4000~K for the \ion{Na}{1} and both
\ion{Fe}{1} features.  The $K$-band EWs in supergiants seem to decrease more
rapidly at $T_{\rm eff} \geq 4500~{\rm K}$ than those in giants but this needs
to be confirmed with additional data given the poor sampling in the range 
$\rm 4500 - 6000~K$ at high luminosity.  The giants generally form a tight
distribution in the EW versus \teff\ diagrams whereas the supergiants
display a larger scatter.  This is presumably due to the larger uncertainties
in the \teff\ calibration and spectral type assignment, and the larger range 
in luminosity within classes I/II \citep[{\em e.g. \rm}][]{Hum84}.

No metallicity effect is obvious in the data presented here, mainly because
of the limited range in [Fe/H] covered by most stars in the samples
and the fairly large uncertainties in [Fe/H] determinations
(see table~\ref{tab-cat}).  From previous empirical
and theoretical studies, a variation of the EWs is expected at more 
extreme metallicities, with an increase from low to high [Fe/H]
({\em e.g.} \citealt{Fro78}; \citealt*{Fro83}; \citealt*{Ter91};
\citealt{Ali95}; \citealt{OFFO97}; \citealt{OO98}; \citealt*{Ste00}).
Part of the dispersion of the data probably reflects the small
differences in [Fe/H] in addition to small variations in intrinsic
\teff\ and other parameters among stars of a given spectral type.

The dwarfs are clearly distinguished by lower EWs for
$T_{\rm eff} \leq 4000-4500~{\rm K}$, except for \ion{Na}{1} and
\ion{Mg}{1}.  The uncertain chemical composition of the coolest
dwarfs from KH86 (especially for the \ion{M2+}{5} star which may
belong to the halo population) prevent further interpretation here.
More details on dwarfs can be found
in \citet{Ali95}, DBJ96, \citet{Ram97}, and \citet{Mey98}. 
The EWs for the first overtone
CO bandheads measured in both carbon stars lie on the loci defined
by the normal red giants but, obviously, no conclusion pertaining
to these particular types can be drawn from such a limited sample.
\citet{Alv00} found from a larger sample that carbon-rich AGB stars
exhibit the full range in CO bandhead strength observed for cool,
oxygen-rich evolved stars, including supergiants.  In the rest of 
this paper, the discussion of EWs and combinations thereof focusses
on normal red giants and supergiants.

The behaviours of the various features with spectral type and luminosity
class seen in figure~\ref{fig-EWs} have been amply discussed in the
literature cited in this work.
It is important to emphasize that the discussion here is intended at
identifying empirical indicators useful for spectral classification and
population synthesis.  A detailed physical interpretation is beyond
the scope of this paper; efforts have been made in that sense by 
{\em e.g.} \citet{McW84}, OMO93, and \citet{Ali95}. 
Such an interpretation is far from trivial because of the contributions
of lines of various species within the bandpasses used to measure the
features and the continuum at moderate resolution.
In addition, other relevant parameters have to be accounted
for, such as the surface gravity, the detailed chemical
composition, and the micro-turbulence velocity.

As an illustration,  figure~\ref{fig-ID} shows the line identifications 
of \citet{WH96} in their $R \geq 45000$ spectra of late-type stars within 
the bandpasses used in this work to quantify the atomic features in the 
$K$-band.  Compared to the molecular features, they are
more ``contaminated,'' and include contributions from \ion{Sc}{1},
\ion{Ti}{1}, \ion{V}{1}, \ion{Si}{1}, \ion{S}{1}, and HF\,(1,0) lines.
In particular, the variations with
temperature and luminosity of the \ion{Na}{1}, \ion{Ca}{1}, and
\ion{Fe}{1} features seem primarily governed by \ion{Sc}{1}, \ion{Ti}{1},
and \ion{V}{1}, which may even become the dominant sources of absorption
in the coolest giants and supergiants.  As noted by \citet{WH96},
the identifications adopted by KH86 are more appropriate for dwarfs than
for giants and supergiants, not surprisingly because they were based on
the comparison of the spectrum of the \ion{K5}{5} star 61\,Cyg\,A with
those of sunspots.

\subsection{Stellar temperature and luminosity indicators}  \label{Sub-indic}

Several empirical diagnostics of spectral type and luminosity
class can be identified from the EWs discussed above.  Due to the
restricted range in metallicities of the stellar samples, these tools
are valid for populations with near-solar abundances.

The \ewcoh\ and \ewca\ constitute pure \teff\ indicators.  None of the
absorption features discussed here varies strictly with luminosity class,
so distinguishing giants from supergiants requires combining 
\teff -sensitive EWs, including at least one also sensitive to 
luminosity class.  The most sensitive luminosity indicator is \ewcoa. 
The \ewcoa/\ewcob\ and \ewcoa/\ewcoc\ ratios plotted in 
figure~\ref{fig-indices} show that all three EWs carry nearly
the same information about the spectral type and luminosity 
class for giants and supergiants (the sharp increase of \ewcoa/\ewcoc\ 
at temperatures above 4500~K may be an artefact due to the 
extreme weakness of the $^{13}$CO\,(2,0) bandhead).  
For the coolest stars ($T_{\rm eff} \lesssim 4000~{\rm K}$) and with
sufficient S/N ratio measurements, \ewna, \ewfea, and \ewfeb\ also 
discriminate between luminosity classes III and I/II.

Various combinations of EWs (sums and ratios) were explored as possible
temperature and luminosity indicators.  In most cases, the composite
indices displayed too large a scatter to provide better diagnostics
than single EWs alone.  However, a few proved to be potentially useful
and are shown in figure~\ref{fig-indices}.  
As demonstrated by OMO93, \cohsi\ and \cohcok\ are sensitive \teff\
indicators, and the former is not affected by a luminosity dependence.
The \ewcoa/\ewmg\ ratio has a pronounced temperature sensitivity, mostly due 
to \ewcoa, but the number of data points is smaller and the scatter is larger
than for \cohsi.  The sum of \ewfea\ and \ewfeb\ provides a higher S/N
ratio measurement than the individual EWs, and a temperature
diagnostic with luminosity sensitivity below 4000~K.  Combining the sharp
increase of \ewcoa\ and \ewca\ with decreasing \teff\ and the luminosity
sensitivity of \ewcoa\ results in a \ewcoa/\ewca\ ratio distinguishing
giants from supergiants at $T_{\rm eff} \lesssim 4500~{\rm K}$.
\citet{Ram97} showed that the ratio of \ewcoa\ and 
$(W_{\rm Ca} + W_{\rm Na})$ constitutes a very good discriminator
between giants and dwarfs below 4800~K, independent of \teff
\footnote{\citeauthor{Ram97}'s EWs are measured differently,
but the conclusions are the same with the definitions adopted here and
the EWs for dwarfs measured on the KH86 spectra.}.  As can be seen in
figure~\ref{fig-indices}, it is however less powerful than \ewcoa/\ewca\ 
for differentiating giants from supergiants because of the luminosity 
sensitivity of \ewna.  Due to similar behaviours with \teff\ and luminosity 
class of \ewcoa\ and \ewna, their ratio is in fact essentially identical in 
giants and supergiants (not shown in figure~\ref{fig-indices}).

\subsection{Contamination by excess continuum sources}  
            \label{Sub-dilution}

The interpretation of the EWs becomes more intricate in objects where
additional continuum sources contribute to the near-infrared emission.
In particular, emission from small interstellar dust grains heated at 
$T \sim 1000~{\rm K}$ by OB stars or by an AGN, photospheric emission
from hot stars, or nebular free-free and free-bound emission may ``dilute''
the absorption features due to the cool stars in star-forming and AGN 
galaxies.  In this case, the properties of the cool stellar population
can be determined from the relative strength of features close enough in
wavelength to be affected by similar amounts of dilution, making the
ratio of their EWs insensitive to such effects.  In addition, the
combination of dilution-free indicators with diluted EWs allows
one to estimate the excess continuum emission without requiring any
knowledge about the properties of the sources, which are usually
very uncertain.

\citet{OOKM95} proposed \cohsi\ as a dilution-free temperature 
indicator, and used it in combination with \ewcoh\ and \cohcok\ 
to constrain the amount of dilution around 1.6\micr\ and 2.3\micr\ 
in various types of galaxies.  For this purpose, they used spectroscopic
equivalents of colour-magnitude plots, namely the 
\ewcoh\ versus \cohsi\ and \ewcoh\ versus \cohcok\ diagrams,
reproduced in figure~\ref{fig-dilution}.  In these diagrams, 
composite stellar populations whose near-infrared emission is 
undiluted occupy regions falling on the observed distributions 
for individual stars.  The amount of dilution near 1.6\micr\ is
determined from the vertical displacement with respect to the locus
of stars in the \ewcoh\ versus \cohsi\ diagram.  Similarly,
the amount of dilution near 2.3\micr\ is estimated from the horizontal
displacement in the \ewcoh\ versus \cohcok\ diagram, once the
\ewcoh\ has been corrected for dilution.  However, because giants and
supergiants lie on separate branches in the latter plot, two solutions
for the 2.3\micr\ dilution are possible unless the luminosity class
can be constrained independently.

Similar diagrams constructed strictly from the $K$-band data are less
satisfactory mainly because of larger scatter and/or lack of strong
enough temperature dependence in ratios of EWs.  Furthermore, the 
variation with luminosity class of almost all indices at low temperatures
may introduce degeneracy in the determination of the amount of dilution. 
The best diagrams in terms of smaller scatter
are \ewca\ versus \ewcoa/\ewmg\ and \ewcoa\ versus \ewcoa/\ewmg, 
also shown in figure~\ref{fig-dilution}.
The \ewcoa/\ewmg\ ratio constitutes the dilution-free temperature 
indicator while \ewca\ and \ewcoa\ are potentially affected by dilution,
which can then be estimated from the vertical displacement relative to
the distribution of stars.
The \ewca\ versus \ewcoa/\ewmg\ diagram would in principle be 
superior because both indices are only sensitive to \teff, yielding
an unambiguous solution for the dilution.  However, the relatively 
loose distribution of the data limits the accuracy.  The \ewcoa\
versus \ewcoa/\ewmg\ diagram exhibits the characteristic double-branch 
behaviour of other similar diagrams due to luminosity effects.
The difference in dilution inferred assuming a population of supergiants 
instead of giants can amount to 50\% for very cool populations.

\subsection{Applications to stellar population studies}  \label{Sub-appl}

Applications of the diagnostic tools discussed in this paper to
studies of stellar clusters and galaxies have shown their usefulness
for constraining the composition of evolved stellar populations
and the amount of excess continuum emission ({\em e.g.}
\citealt{OOKM95}; \citealt*{Bok97}; \citealt{Tha97}; \citealt{Oli99};
\citealt{FS00a}).  Implemented in 
evolutionary synthesis models, they can also provide quantitative
constraints on the star formation history ({\em e.g.}
\citealt{Ori00, FS00b}).

In applying these tools, it is however important to bear in mind 
some of their potential limitations, briefly outlined here.
The frequently limited S/N ratio for extragalactic sources may
hamper reliable measurements of weak features.
The dispersion in the template stellar data alone
limits the accuracy in determinations of \teff\ and of the dilution.
This depends on the chosen diagnostics as well as the actual
\teff\ range and luminosity class of the stars under consideration,
but typical uncertainties are of about $\rm \pm 250~K$ for \teff\
(or about $\pm 3$ spectral sub-classes within K- and M-types)
and of $\approx 10\% - 20\%$ for the dilution.
Degeneracy between luminosity class and dilution may lead to two possible
solutions for these properties.  The interpretation of the feature
strengths is further complicated when distinct cool populations contribute
to the integrated emission, such as red supergiants formed in a recent
starburst and older, preexisting giants.
Finally, predictions of the variations with time of the EWs using
evolutionary synthesis models can be affected by additional 
uncertainties which arise notably from the possible inadequacy of
current stellar tracks to represent the evolution of intermediate-mass 
stars along the AGB \citep[{\em e.g. \rm}][]{Ori00} --- not mentioning
the scarcity of data for AGB stars.

The various indicators remain nonetheless valuable tools for extragalactic
applications as specific probes of the cool, evolved stars.
In cases where their diagnostic power is limited, they
benefit to be complemented with additional constraints also sensitive
to the stellar contents such as the mass-to-light ratio, or with
the spatial distribution of the EWs and of the stellar near-infrared 
continuum emission which can provide indirect evidence concerning the
nature of the stellar population at a given location
({\em e.g.} \citealt{Oli99, FS00a}).

\section{Summary}  \label{Sect-conclu}

$K$-band spectra at $R \approx 830$ and 2000 of 31 late-type giants 
and supergiants mostly with near-solar abundances, and of two carbon stars 
were obtained with the aim of widening existing near-infrared stellar 
libraries useful for population synthesis of clusters and galaxies.
The EWs of several of the most prominent absorption features were computed:
the \ion{Na}{1} 2.2076\micr\ and \ion{Ca}{1} 2.2636\micr\ features,
and the $^{12}$CO\,(2,0), $^{12}$CO\,(3,1), and $^{13}$CO\,(2,0) bandheads.
Relationships between these EWs and the instrumental resolution
were derived.  These allow comparison between data sets obtained with
different resolving powers in the range $R \sim 600-3000$ or, equivalently,
affected by velocity broadening up to $\rm \sigma = 200~km\,s^{-1}$.
In addition, the EWs of the \ion{Mg}{1} 2.2814\micr, \ion{Fe}{1} 2.2263, and
\ion{Fe}{1} 2.2387\micr\ features were measured in the higher resolution 
spectra.

The 3D data set was augmented with EWs obtained from other similar 
stellar libraries, and complemented with EWs of the $^{12}$CO\,(6,3)
bandhead and \ion{Si}{1} 1.5892\micr\ feature in the $H$-band.
This extended database was used to investigate spectroscopic
diagnostics of stellar spectral type and luminosity class
as well as of the contribution by excess continuum sources. 
The main conclusions are the following. 

$\bullet$ The various data sets compare remarkably well.  Compiled together,
they better constrain the variations of the absorption features with 
spectral type and luminosity class found previously by various authors
based on smaller samples. 

$\bullet$ Several features provide useful tools in population
studies of stellar clusters and galaxies.  Among the EWs considered,
\ewcoa, \ewcoh\ and \ewsi\ first proposed by OMO93 remain the
best set of indicators for constraining the average spectral type
and luminosity class of cool evolved stars, and the amount of dilution
near 1.6\micr\ and 2.3\micr.  The \ewca\ and \ewmg\ offer alternatives
to \ewcoh\ and \ewsi\ in the $K$-band. 

$\bullet$ There is still an important need for empirical and theoretical 
work to further investigate metallicity effects and better constrain the 
stellar properties and evolution at high luminosities, including
red supergiants and AGB stars.
The work initiated by Origlia, Oliva and coworkers \citep{OFFO97, OO98, Ori00}
and by Lan\c{c}on and her collaborators \citep{Lan98, Lan99, Alv00}
constitute valuable contributions in this direction.

\acknowledgments

I am grateful to the MPE-3D team for help with the observations.
I wish to thank E. Oliva, L. Tacconi-Garman, L. Tacconi, R. Maiolino,
and R. Genzel for interesting discussions and useful comments on
the manuscript.  I also thank the anonymous referee for further 
helpful comments and suggestions.  This research made use of
the SIMBAD database operated by CDS in Strasbourg, France.
I acknowledge the Fonds pour les Chercheurs et l'Aide \`a la Recherche
(Gouvernement du Qu\'ebec, Canada) for a Graduate Scholarship, and the
Max-Planck-Institut f\"ur extraterrestrische Physik as well as the Service
d'Astrophysique of the CEA Saclay for additional financial support.

\clearpage

\figcaption[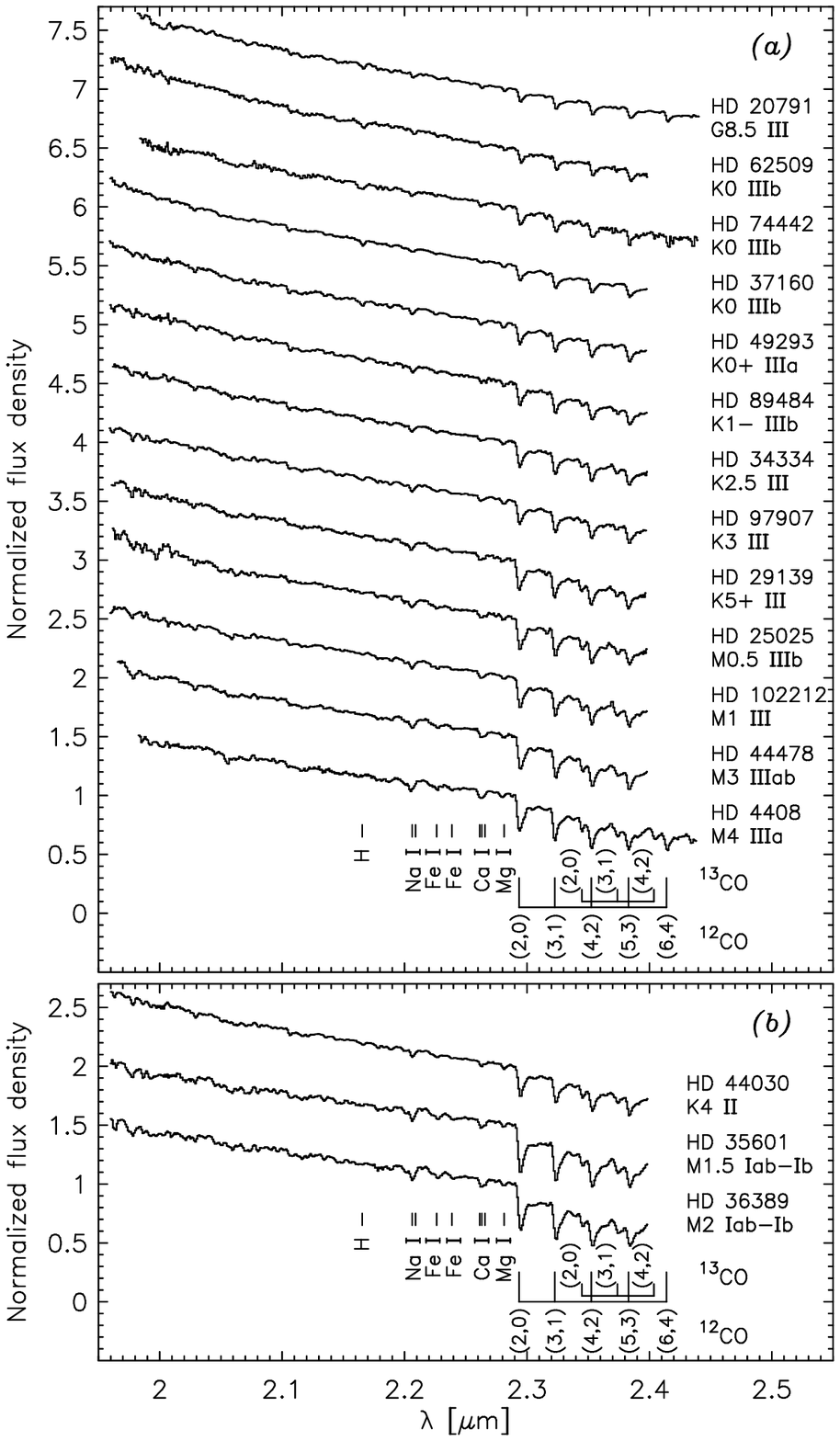]
{
3D spectra at $R \approx 830$:
{\em (a)} giants, and {\em (b)} supergiants.
The spectra are normalized to unity between 2.2875\micr\ and 2.2910\micr.
In each panel, the vertical axis is the appropriate flux scale for the
bottom spectrum and each successive spectrum is shifted upwards by 0.5.
The strongest features longwards of 2.15\micr\ are labeled, following
the nomenclature of \protect\citet{KH86}.
\label{fig-SpecLR}
}

\figcaption[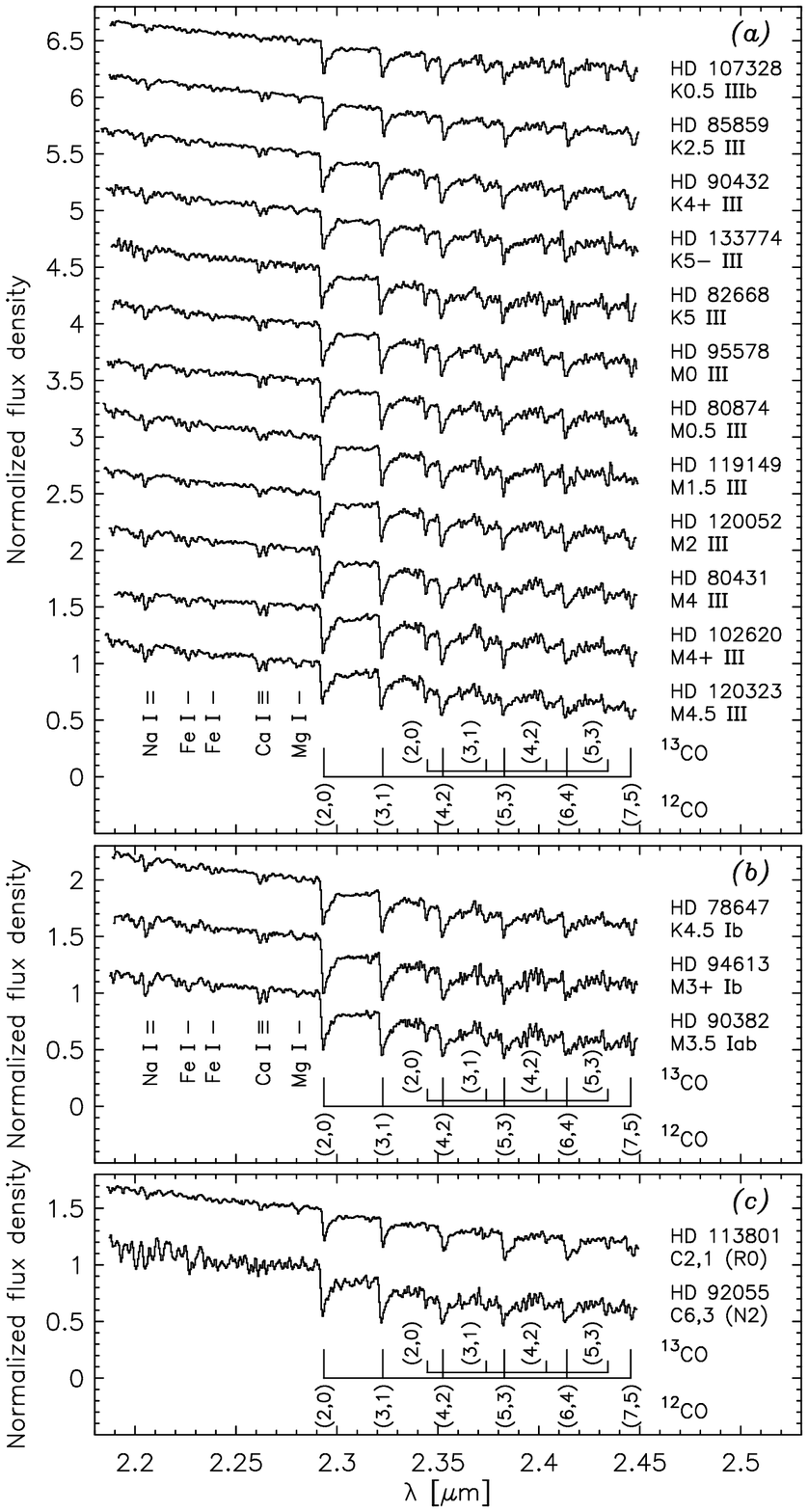]
{
3D spectra at $R \approx 2000$:
{\em (a)} giants, {\em (b)} supergiants, and {\em (c)} carbon stars.
The spectra are normalized to unity between 2.2875\micr\ and 2.2910\micr.
In each panel, the vertical axis is the appropriate flux scale for the
bottom spectrum and each successive spectrum is shifted upwards by 0.5.
The strongest features longwards of 2.15\micr\ are labeled, following
the nomenclature of \protect\citet{KH86}.
\label{fig-SpecHR}
}

\figcaption[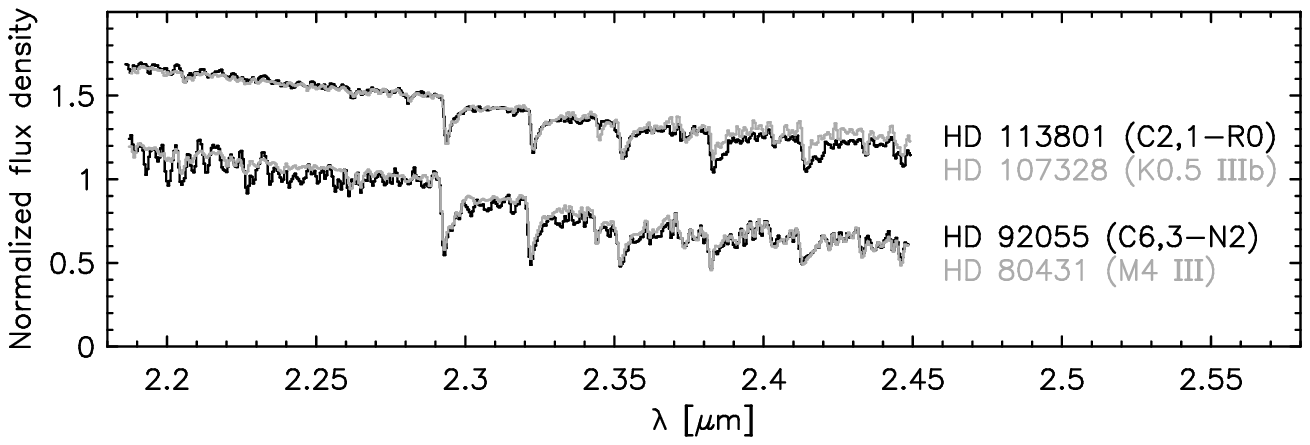]
{
Comparison of the spectra of the carbon stars (black lines) with those
of normal giants of similar effective temperature (grey lines).
The spectra are normalized and displayed on a linear flux scale
as for figure~\ref{fig-SpecLR}.
\label{fig-C-III}
}

\figcaption[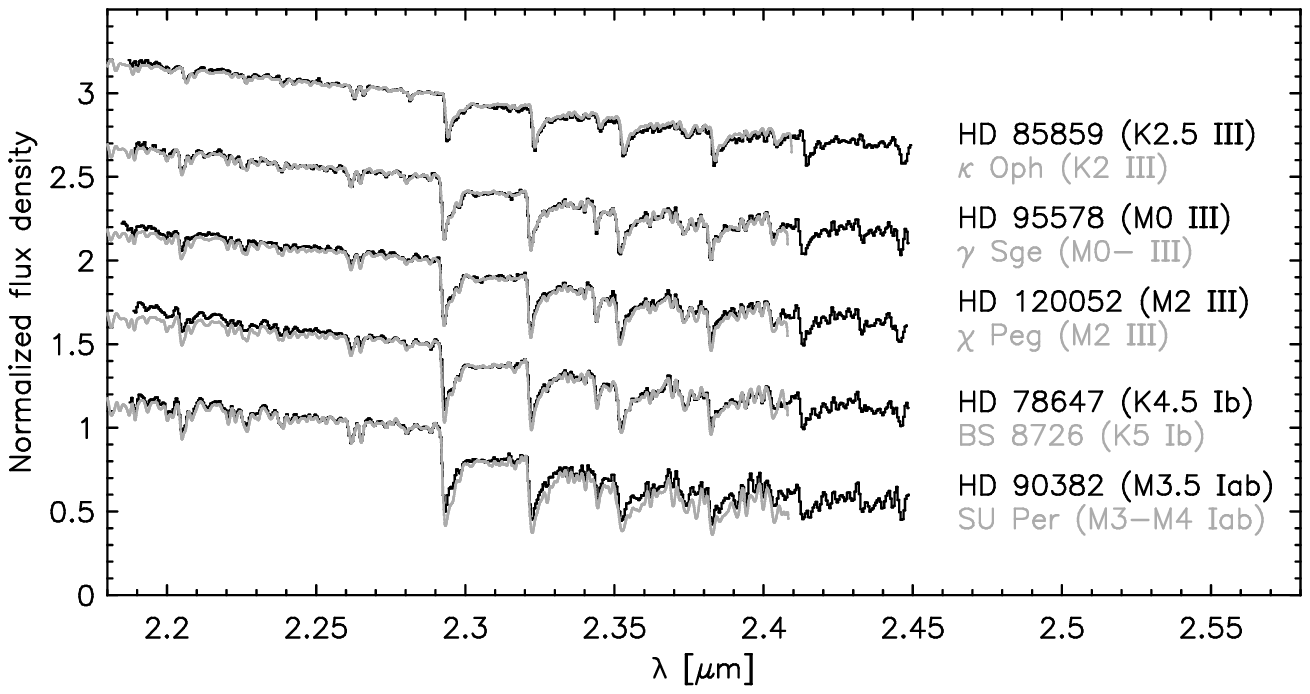]
{ 
Comparison of a subset of 3D spectra at $R \approx 2000$ (black lines) 
with spectra of stars having similar spectral types from \protect\citet{KH86},
convolved to the resolution of the 3D data (grey lines).
The latter have been multiplied by a power law continuum proportional
to $\lambda^{-3.94}$ in order to recover the original energy distribution
of the stars (see section~\ref{Sub-spec}).  All spectra are normalized and 
displayed on a linear flux scale as for figure~\ref{fig-SpecLR}.
\label{fig-3D-KH}
}

\figcaption[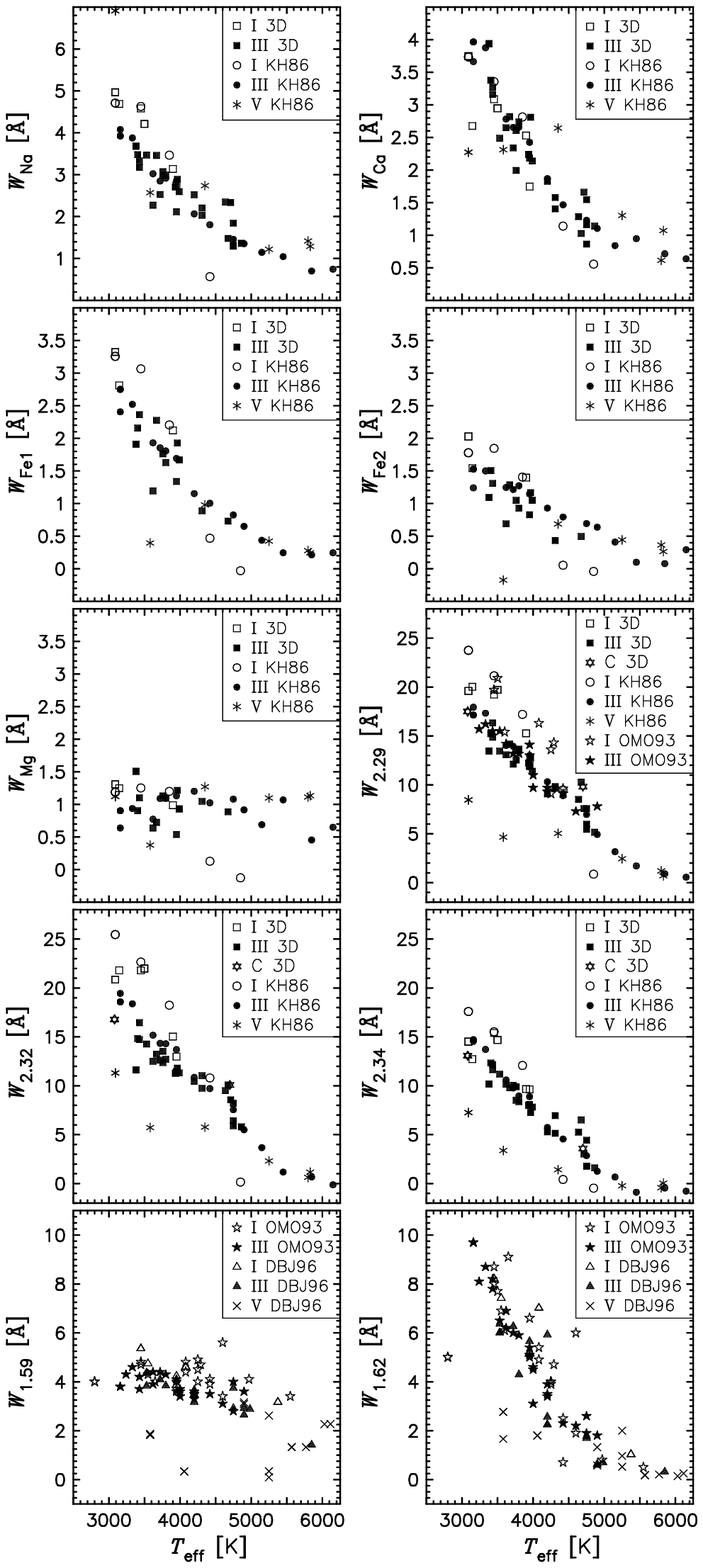] 
{ 
Variations with stellar effective temperature of the equivalent widths 
of the atomic and molecular absorption features discussed in the text.
Open symbols represent data for supergiants and carbon stars, filled
symbols represent data for giants, and crosses represent data for dwarfs.
Different symbols are also used to distinguish between data from various
authors, as indicated in the insets (3D: this work; 
KH86: \protect\citealt{KH86}; OMO93: \protect\citealt{OMO93}; 
DBJ96: \protect\citealt{DBJ96}).
The typical measurement uncertainties are $\pm 0.5$~\AA\ for the atomic
features and \ewcoh, $\pm 0.8$~\AA\ for \ewcoa, and $\pm 1.0$~\AA\ for \ewcob\
and \ewcoc.
\label{fig-EWs}
}

\figcaption[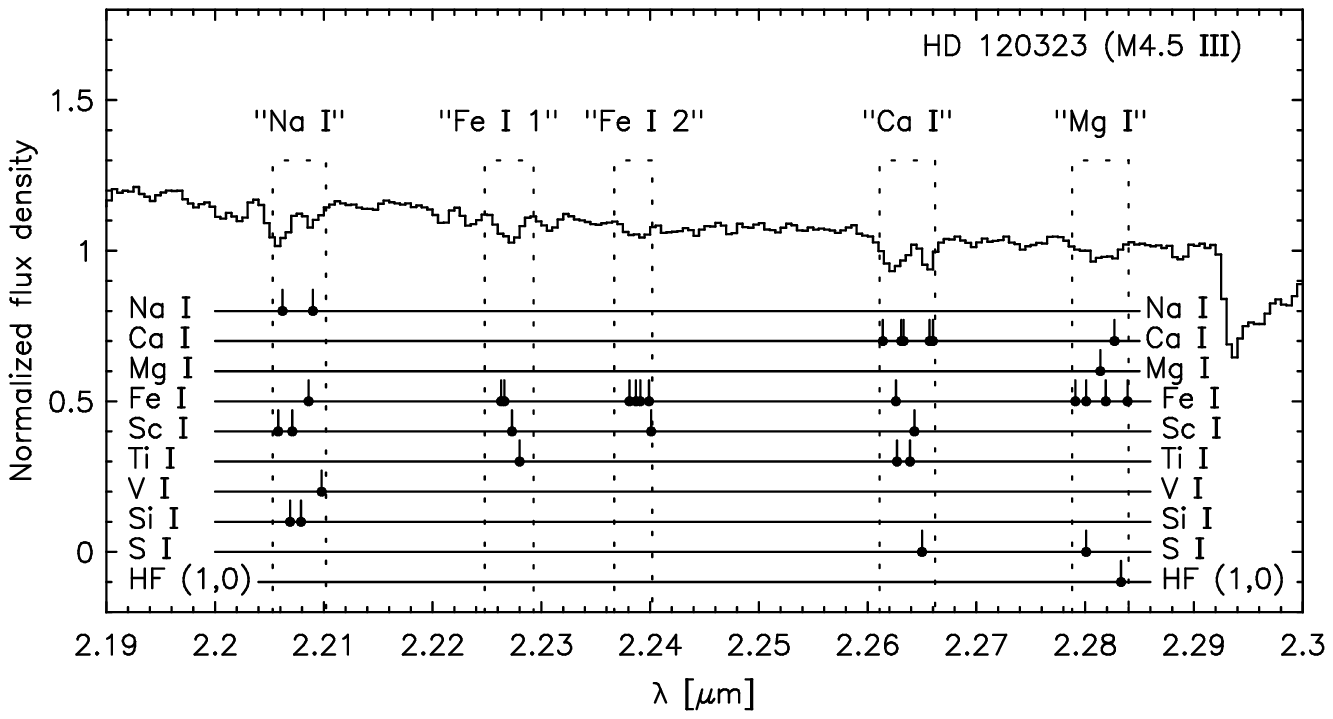]
{
Positions of the lines of various species contributing to the
$K$-band atomic features considered in this work, as identified 
by \protect\citet{WH96} in $R \geq 45000$ spectra of red dwarfs, giants,
and supergiants.
The bandpasses used to integrate the depth of the atomic features
are indicated on the spectrum of HD\,120323 by the dotted lines
(see table~\ref{tab-Def}).
\label{fig-ID}
}

\figcaption[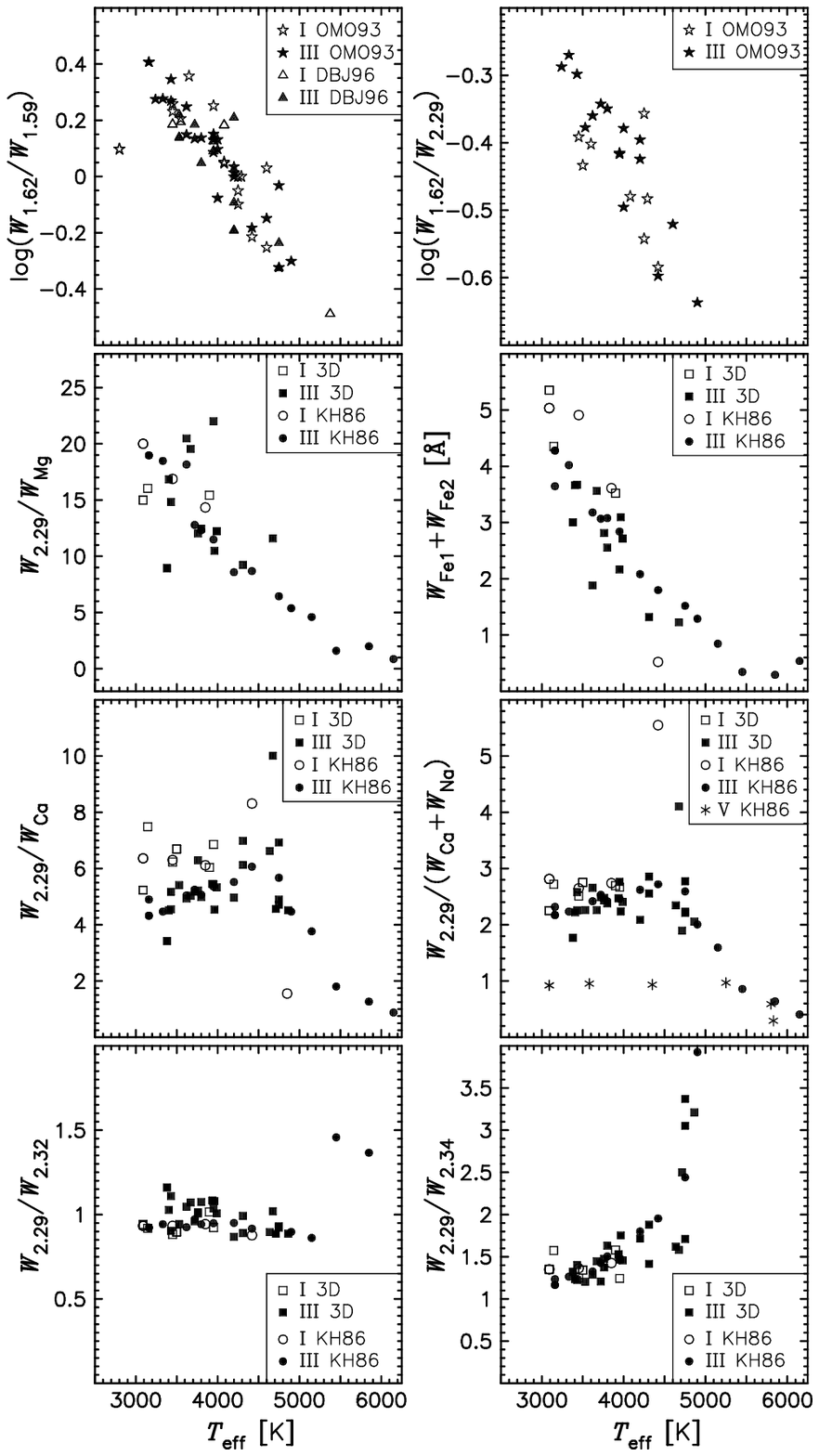]
{
Variations of composite spectroscopic indices with stellar effective 
temperature. 
The symbols are the same as for figure~\ref{fig-EWs}, and labeled in the
insets.  Data of dwarfs are shown for $W_{2.29}/(W_{\rm Ca} + W_{\rm Na})$
to illustrate that this indicator is a less powerful discriminator between
giants and supergiants than it is between dwarfs and giants 
(see section~\ref{Sub-indic}).
The sharp increase of \ewcoa/\ewcoc\ at high \teff\ likely is an
artefact reflecting the extreme weakness of the $^{13}$CO\,(2,0) bandhead.
\label{fig-indices}
}

\pagebreak

\figcaption[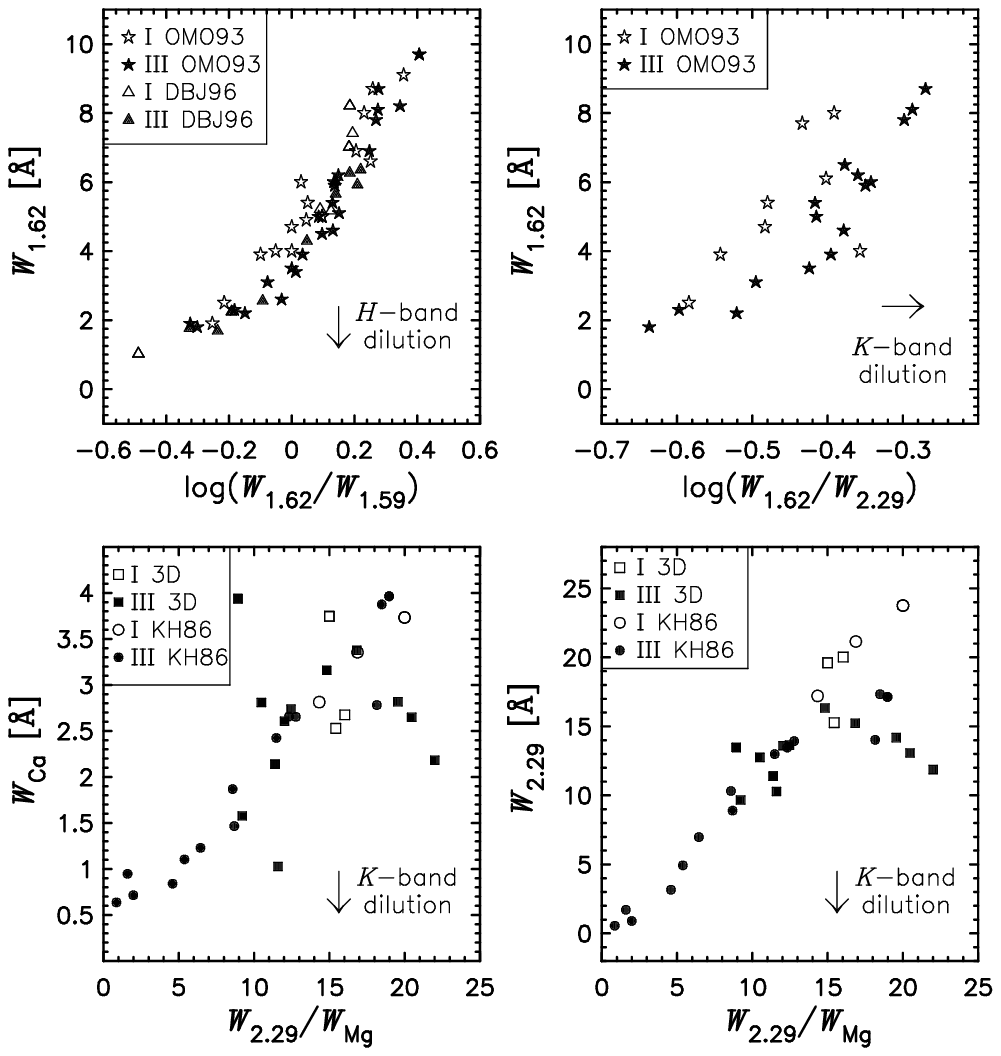]
{
Diagnostic diagrams for estimating the amount of dilution of the 
stellar absorption features by excess continuum emission.
The composite indices used for the horizontal axes are sensitive to \teff\ 
only.  The arrows show the effects on the indices of dilution near 1.6\micr\ 
(top left panel) and near 2.3\micr\ (other panels).
The symbols are the same as for figure~\ref{fig-EWs}, and labeled in the
insets.  
\label{fig-dilution}
}

\clearpage

\begin{deluxetable}{lllll}
\tabletypesize{\small}
\tablecolumns{5}
\tablewidth{0pt}
\tablenum{1}
\tablecaption{Program stars \label{tab-cat}}
\tablehead{
\colhead{Star} & \colhead{Spectral type} & 
\colhead{References} & \colhead{[Fe/H]\,\tablenotemark{a}} & 
\colhead{References}
}
\startdata
HD 44030 & K4 II & 1 & $-0.22$; $-0.7$ & 4; 1 \\
HD 78647 & K4.5 Ib & 2 & 0.23 & 1 \\
HD 35601 & M1.5 Iab-Ib & 2 & $-0.24$ & 1 \\
HD 36389 & M2 Iab-Ib & 2 & 0.11 & 1 \\
HD 94613 & M3+ Ib & 2 & ... & ... \\ 
HD 90382 & M3.5 Iab & 2 & ... & ... \\ 
HD 20791 & G8.5 III & 2 & ... & ... \\
HD 62509 & K0 IIIb & 2 & 0.01; $-0.07$; $-0.51$ to 0.16 & 4; 5; 1 \\
HD 74442 & K0 IIIb & 2 & $-0.11$; $-0.13$ & 4; 5 \\
HD 37160 & K0 IIIb Fe-2 & 2 & $-0.58$; $-0.63$; $-0.70$ to $-0.19$ & 4; 5; 1 \\
HD 49293 & K0+ IIIa & 2 & $-0.08$; $-0.12$ & 4; 5 \\
HD 107328 & K0.5 IIIb Fe-0.5 & 2 & $-0.32$; $-0.48$; $-0.64$ to $-0.12$ & 4; 5; 1 \\
HD 89484 & K1$-$ IIIb Fe-0.5 & 2 & $-0.25$; $-0.49$; $-0.41$ to 0.09 & 4; 5; 1 \\
HD 85859 & K2.5 III & 2 & $-0.08$; $-0.03$ & 4; 5 \\
HD 34334 & K2.5 III Fe-1 & 2 & $-0.26$; $-0.46$; $-0.39$, $-0.25$ & 4; 5; 1 \\
HD 97907 & K3 III & 1 & $-0.09$ & 4 \\
HD 90432 & K4+ III & 2 & $-0.24$; $-0.12$ & 4; 5 \\
HD 133774 & K5$-$ III & 2 & $-0.41$; 0.01 & 4; 5 \\
HD 82668 & K5 III & 2 & ... & ... \\
HD 29139 & K5+ III & 2 & $-0.16$; $-0.34$; $-0.33$ to 0.00 & 4; 5; 1 \\
HD 95578 & M0 III & 2 & ... & ... \\
HD 80874 & M0.5 III & 2 & ... & ... \\
HD 25025 & M0.5 IIIb Ca-1 & 2 & ... & ... \\
HD 102212 & M1 III & 2 & $-0.09$ & 1\\
HD 119149 & M1.5 III & 2 & ... & ... \\
HD 120052 & M2 III & 2 & ... & ... \\
HD 44478 & M3 IIIab & 2 & $-0.05$ & 1 \\
HD 80431 & M4 III & 2 & ... & ... \\
HD 4408 & M4 IIIa & 2 & ... & ... \\
HD 102620 & M4+ III & 2 & ... & ... \\ 
HD 120323 & M4.5 III & 2 & ... & ... \\ 
HD 113801 & C2,1 (R0) & 3 & $-0.26$ & 1 \\ 
HD 92055 & C6,3 (N2) & 3 & $-0.1$ & 1 \\ 
\enddata
\tablerefs
{
1: \citet{Cay92}; 2: \citet{KM89}; 3: \citet{Yam72}; 4: \citet{Tay91};
5: \citet{McW90}.
}
\tablenotetext{a}
{
Metallicities are averages computed by \citet{Tay91} from data in the
literature, values collected from the literature compiled by
\citet{Cay92}, and best fits to high resolution optical spectra
and photometric data using model atmospheres by \citet{McW90}.
}
\end{deluxetable}

\clearpage

\begin{deluxetable}{lclcccc}
\tabletypesize{\small}
\tablecolumns{7}
\tablewidth{0pt}
\tablenum{2}
\tablecaption{Log of the observations \label{tab-Obs}}
\tablehead{
\colhead{Star} & \colhead{$t_{\rm int}$\,\tablenotemark{a}} &
\colhead{Calibrator} & \colhead{Run\,\tablenotemark{b}} & 
\colhead{Range} & \colhead{$R$} & 
\colhead{est. S/N\,\tablenotemark{c}} \\
 & \colhead{(s)} & & & \colhead{($\rm \mu m$)} & & 
}
\startdata
HD 44030  & 100  & PPM 95062 (F8 V) & CA95 &
$1.90-2.40$  & 830  & 376 \\ 
HD 78647  & 10  & BS 3699 (A8 Ib) & LS96 &
$2.18-2.45$  & 2000  & 113 \\
HD 35601  & 54  & PPM 95062 (F8 V) & CA95 &
$1.90-2.40$  & 830  & 205 \\
HD 36389  & 4.5  & PPM 95062 (F8 V) & CA95 &
$1.90-2.40$  & 830  & 213 \\
HD 94613  & 38  & HD 87504 (B9 III-IV) & LS96 &
$2.18-2.45$  & 2000  & 99 \\
HD 90382  & 40  & BS 4037 (B8 IIIe) & LS96 &
$2.18-2.45$  & 2000  & 85 \\
HD 20791  & 112  & BS 1153 (B3 V) & LP96 &
$1.90-2.40$  & 830  & 328 \\
HD 62509  & 2.7  & BS 2890 (A2 V) & CA95 &
$1.90-2.40$  & 830  & 218 \\
HD 74442  & 14.5  & BS 3510 (G1 V) & LP96 &
$1.90-2.40$  & 830  & 309 \\
HD 37160  & 95  & PPM 95062 (F8 V) & CA95 &
$1.90-2.40$  & 830  & 343 \\
HD 49293  & 36  & BS 2779 (F8 V) & CA95 &
$1.90-2.40$  & 830  & 277 \\
HD 107328  & 40  & BS 4757 (B9.5 V) & LS96 &
$2.18-2.45$  & 2000  & 112 \\
HD 89484  & 49  & BS 4039 (F8 V) & CA95 &
$1.90-2.40$  & 830  & 212 \\
HD 85859  & 95  & HD 71459 (B3 V) & LS96 &
$2.18-2.45$  & 2000  & 108 \\
HD 34334  & 58  & PPM 95062 (F8 V) & CA95 &
$1.90-2.40$  & 830  & 349 \\
HD 97907  & 140  & PPM 158259 (F8 V) & CA95 &
$1.90-2.40$  & 830  & 418 \\
HD 90432  & 38  & HD 87504 (B9 III-IV) & LS96 &
$2.18-2.45$  & 2000  & 155 \\
HD 133774  & 20  & HD 136664 (B4 V) & LS96 &
$2.18-2.45$  & 2000  & 107 \\
HD 82668  & 9.5  & HD 87504 (B9 III-IV) & LS96 &
$2.18-2.45$  & 2000  & 83 \\
HD 29139  & 8.7  & BS 788 (F9 V) & CA95 &
$1.90-2.40$  & 830  & 323 \\
HD 95578  & 9  & HD 87504 (B9 III-IV) & LS96 &
$2.18-2.45$  & 2000  & 119 \\
HD 80874  & 19  & HD 71459 (B3 V) & LS96 &
$2.18-2.45$  & 2000  & 112 \\
HD 25025  & 14.5  & BS 788 (F9 V) & CA95 &
$1.90-2.40$  & 830  & 283 \\
HD 102212  & 49  & PPM 158259 (F8 V) & CA95 &
$1.90-2.40$  & 830  & 328 \\
HD 119149  & 40  & HD 129116 (B3 V) & LS96 &
$2.18-2.45$  & 2000  & 100 \\
HD 120052  & 19  & HD 129116 (B3 V) & LS96 &
$2.18-2.45$  & 2000  & 132 \\
      &       &  BS 4134 (F6 V) &     &
      &       &       \\
HD 44478  & 2.7  & PPM 95062 (F8 V) & CA95 &
$1.90-2.40$  & 830  & 344 \\
HD 80431  & 100  & BS 3836 (A5 IV-V) & LS96 &
$2.18-2.45$  & 2000  & 142 \\
HD 4408  & 22.2  & HD 4676 (F8 V) & LP96 &
$1.90-2.40$  & 830  & 290 \\
HD 102620  & 6  & BS 4662 (B8 IIIpHgMn) & LS96 &
$2.18-2.45$  & 2000  & 133 \\
HD 120323  & 9.5  & BS 4979 (G3 V) & LS96 &
$2.18-2.45$  & 2000  & 99 \\ 
HD 113801  & 600  & BS 5210 (B5 III) & LS96 &
$2.18-2.45$  & 2000  & 82 \\
HD 92055  & 10  & BS 3836 (A5 IV-V) & LS96 &
$2.18-2.45$  & 2000  & 25 \\ 
\enddata
\tablenotetext{a}
{
Total on-source integration time per wavelength channel.
}
\tablenotetext{b}
{
CA95 : January 1995 run at the 3.5~m telescope at Calar Alto, Spain;
LS96 : April 1996 run at the ESO 2.2~m telescope at La Silla, Chile.
LP96 : January 1996 run at the 4.2~m William-Herschel-Telescope 
on the Canary Islands, Spain;
}
\tablenotetext{c}
{
Estimated signal-to-noise ratio per wavelength channel evaluated
on the reduced spectra between 2.242\micr\ and 2.258\micr\
(see section~\ref{Sub-datared}).
}
\end{deluxetable}

\clearpage

\begin{deluxetable}{llll}
\tablecolumns{4}
\tablewidth{0pt}
\tablenum{3}
\tablecaption{Continuum points and bandpass edges for the absorption features 
              \label{tab-Def}}
\tablehead{
\colhead{Feature} & \colhead{Symbol} & 
\colhead{Continuum points\,\tablenotemark{a}} &
\colhead{Integration limits} \\ 
      &       & \colhead{($\rm \mu m$)} & \colhead{($\rm \mu m$)} 
}
\startdata
\ion{Si}{1} \ 1.5892\micr  & \,\ \ewsi  & 
1.5850, 1.5930 & \hspace{0.2em} $1.5870-1.5910$ \\
$^{12}$CO (6,3) 1.6187\micr  & \,\ \ewcoh  & 
1.6160, 1.6270 & \hspace{0.2em} $1.6175-1.6220$ \\
\ion{Na}{1} \ 2.2076\micr  & \,\ \ewna  & 
2.1965, 2.2125, 2.2175 & \hspace{0.2em} $2.2053-2.2101$ \\
\ion{Fe}{1} \ 2.2263\micr  & \,\ \ewfea  & 
2.2125, 2.2175, 2.2323 & \hspace{0.2em} $2.2248-2.2293$ \\
\ion{Fe}{1} \ 2.2387\micr  & \,\ \ewfeb  & 
2.2323, 2.2510, 2.2580 & \hspace{0.2em} $2.2367-2.2402$ \\
\ion{Ca}{1} \ 2.2636\micr  & \,\ \ewca  & 
2.2510, 2.2580, 2.2705, 2.2760  & \hspace{0.2em} $2.2611-2.2662$ \\
\ion{Mg}{1} \ 2.2814\micr  & \,\ \ewmg  & 
2.2705, 2.2760, 2.2900 & \hspace{0.2em} $2.2788-2.2840$ \\
$^{12}$CO (2,0) 2.2935\micr  & \,\ \ewcoa  & 
2.2900  & \hspace{0.2em} $2.2924-2.2977$ \\
$^{12}$CO (3,1) 2.3227\micr  & \,\ \ewcob  & 
2.2125, 2.2175, 2.2335, 2.2580, 2.2705, 2.2900  & \hspace{0.2em} $2.3218-2.3272$ \\ 
$^{13}$CO (2,0) 2.3448\micr  & \,\ \ewcoc  & 
(same as for \ewcob) & \hspace{0.2em} $2.3436-2.3491$ \\ 
\enddata
\tablenotetext{a}
{
Central wavelength of the 
$\rm 0.002 - 0.003~\mu m$ wide intervals
used to fit the normalizing continuum.
}
\end{deluxetable}

\clearpage

\begin{deluxetable}{lllllllrrr}
\tabletypesize{\small}
\tablecolumns{10}
\tablewidth{0pt}
\tablenum{4}
\tablecaption{Measured equivalent widths   \label{tab-EWs}}
\tablehead{
\colhead{Star} & \colhead{Type} & \colhead{\ewna} & \colhead{\ewfea} &
\colhead{\ewfeb} & \colhead{\ewca} & \colhead{\ewmg} &
\colhead{\ewcoa} & \colhead{\ewcob} & \colhead{\ewcoc} 
}
\startdata
HD 44030  & K4 II & 2.7 & ... & ... & 1.7 & ... & 12.0 & 13.0 & 9.6 \\
HD 78647  & K4.5 Ib & 3.1 & 2.1 & 1.4 & 2.5 & 1.0 & 15.3 & 15.0 & 9.7 \\
HD 35601  & M1.5 Iab-Ib & 4.2 & ... & ... & 2.9 & ... & 19.7 & 22.0 & 14.7 \\
HD 36389  & M2 Iab-Ib & 4.6 & ... & ... & 3.1 & ... & 19.2 & 21.8 & 15.4 \\
HD 94613  & M3+ Ib & 4.7 & 2.8 & 1.5 & 2.7 & 1.2 & 20.0 & 21.8 & 12.7 \\ 
HD 90382  & M3.5 Iab & 5.0 & 3.3 & 2.0 & 3.7 & 1.3 & 19.6 & 20.9 & 14.5 \\
HD 20791  & G8.5 III & 1.4 & ... & ... & 1.1 & ... & 5.1 & 5.8 & 1.6 \\
HD 62509  & K0 IIIb & 1.3 & ... & ... & 1.1 & ... & 5.5 & 5.9 & 1.8 \\
HD 74442  & K0 IIIb & 1.8 & ... & ... & 1.5 & ... & 7.6 & 8.2 & 4.4 \\
HD 37160  & K0 IIIb Fe-2 & 1.3 & ... & ... & 0.9 & ... & 6.0 & 6.4 & 1.8 \\
HD 49293  & K0+ IIIa & 2.3 & ... & ... & 1.7 & ... & 7.8 & 8.6 & 3.0 \\
HD 107328 & K0.5 IIIb Fe-0.5 & 1.5 & 0.7 & 0.5 & 1.0 & 0.9 & 10.3 & 10.1 & 6.5 \\
HD 89484  & K1$-$ IIIb Fe-0.5 & 2.3 & ... & ... & 1.3 & ... & 8.5 & 9.5 & 5.3 \\
HD 85859  & K2.5 III & 2.2 & 0.9 & 0.4 & 1.6 & 1.0 & 9.7 & 9.7 & 5.1 \\
HD 34334  & K2.5 III Fe-1 & 2.0 & ... & ... & 1.4 & ... & 9.8 & 11.0 & 6.9 \\
HD 97907  & K3 III & 2.5 & ... & ... & 1.8 & ... & 9.1 & 10.5 & 5.3 \\
HD 90432  & K4+ III & 2.6 & 1.7 & 1.0 & 2.1 & 0.9 & 11.4 & 11.3 & 7.8 \\
HD 133774 & K5$-$ III & 2.9 & 1.9 & 1.2 & 2.8 & 1.2 & 12.8 & 11.8 & 7.3 \\
HD 82668  & K5 III & 2.1 & 1.3 & 0.8 & 2.2 & 0.5 & 11.9 & 11.5 & 8.1 \\
HD 29139  & K5+ III & 2.7 & ... & ... & 2.2 & ... & 12.2 & 11.3 & 8.0 \\
HD 95578  & M0 III & 3.0 & 1.6 & 0.9 & 2.7 & 1.1 & 13.6 & 12.7 & 8.4 \\
HD 80874  & M0.5 III & 3.0 & 1.8 & 1.1 & 2.6 & 1.1 & 13.6 & 13.5 & 9.9 \\
HD 25025  & M0.5 IIIb Ca-1 & 3.1 & ... & ... & 2.0 & ... & 12.5 & 12.4 & 8.5 \\
HD 102212 & M1 III & 2.5 & ... & ... & 2.3 & ... & 12.1 & 12.6 & 10.0 \\
HD 119149 & M1.5 III & 3.5 & 2.3 & 1.3 & 2.8 & 0.7 & 14.2 & 13.2 & 9.8 \\
HD 120052 & M2 III & 2.3 & 1.2 & 0.7 & 2.6 & 0.6 & 13.1 & 12.5 & 10.2 \\
HD 44478  & M3 IIIab & 3.5 & ... & ... & 2.5 & ... & 13.5 & 14.3 & 11.2 \\
HD 80431  & M4 III & 3.2 & 2.4 & 1.3 & 3.2 & 1.1 & 16.3 & 14.7 & 11.7 \\
HD 4408   & M4 IIIa & 3.3 & ... & ... & 3.3 & ... & 14.9 & 16.5 & 12.2 \\
HD 102620 & M4+ III & 3.5 & 2.2 & 1.5 & 3.4 & 0.9 & 15.2 & 14.8 & 12.3 \\ 
HD 120323 & M4.5 III & 3.7 & 1.9 & 1.1 & 3.9 & 1.5 & 13.5 & 11.6 & 10.2 \\ 
HD 113801 & C2,1 (R0) & ... & ... & ... & ... & ... & 9.8 & 10.1 & 3.6 \\ 
HD 92055  & C6,3 (N2) & ... & ... & ... & ... & ... & 17.5 & 16.8 & 13.1 \\ 
\enddata
\tablecomments{
The equivalent widths are given in \AA.
The typical measurement uncertainties are $\pm 0.5$~\AA\ 
for the atomic features, $\pm 0.8$~\AA\ for \ewcoa,
and $\pm 1.0$~\AA\ for \ewcob\ and \ewcoc.
}
\end{deluxetable}


\setcounter{figure}{0}

\clearpage

\begin{figure}[p]
\epsscale{1.0}
\plotone{Forster-Schreiber.fig1.ps}
\vspace{-2.0cm}
\figcaption[]{}
\end{figure}

\clearpage

\begin{figure}[p]
\epsscale{1.0}
\plotone{Forster-Schreiber.fig2.ps}
\vspace{-1.0cm}
\figcaption[]{}
\end{figure}

\clearpage

\begin{figure}[p]
\epsscale{1.00}
\plotone{Forster-Schreiber.fig3.ps}
\vspace{-8.0cm}
\figcaption[]{}
\end{figure}

\clearpage

\begin{figure}[p]
\epsscale{1.00}
\plotone{Forster-Schreiber.fig4.ps}
\vspace{-7.0cm}
\figcaption[]{}
\end{figure}

\clearpage

\begin{figure}[p]
\epsscale{1.0}
\plotone{Forster-Schreiber.fig5.ps}
\vspace{-3.5cm}
\figcaption[]{}
\end{figure}

\clearpage

\begin{figure}[p]
\epsscale{1.00}
\plotone{Forster-Schreiber.fig6.ps}
\vspace{-7.0cm}
\figcaption[]{}
\end{figure}

\clearpage

\begin{figure}[p]
\epsscale{1.0}
\plotone{Forster-Schreiber.fig7.ps}
\vspace{-5.0cm}
\figcaption[]{}
\end{figure}

\clearpage

\begin{figure}[p]
\epsscale{1.0}
\plotone{Forster-Schreiber.fig8.ps}
\vspace{-6.0cm}
\figcaption[]{}
\end{figure}

\end{document}